\documentclass[twocolumn,aps,superscriptaddress,amsmath,amssymb]{revtex4}

\usepackage{bm} 
\usepackage{longtable}
\usepackage{tabularx}
\usepackage{adjustbox}
\usepackage[normalem]{ulem}
\usepackage[utf8]{inputenc}

\usepackage{amsfonts}
\usepackage{hyperref}		
\hypersetup{
  colorlinks   = true,		
  urlcolor     = blue,		
  linkcolor    = blue,		
  citecolor   = blue		
}

\usepackage{xcolor}

\usepackage{multirow}

\usepackage{graphicx}

\usepackage{braket}

\begin{document}

\title{Multiple Interacting Photonic Modes in Strongly Coupled Organic Microcavities}

\author{Felipe Herrera}
\affiliation{Department of Physics, Universidad de Santiago de Chile, Av. Victor Jara 3493, Santiago, Chile}

\author{William L. Barnes}
\affiliation{Department of Physics and Astronomy, Stocker Road, University of Exeter, Devon EX4 4QL, United Kingdom}


\begin{abstract}
    Room temperature cavity quantum electrodynamics with molecular materials in optical cavities offers exciting prospects for controlling electronic, nuclear and photonic degrees of freedom for applications in physics, chemistry and materials science. However, achieving strong coupling with molecular ensembles typically requires high molecular densities and substantial electromagnetic field confinement. These conditions usually involve a significant degree of molecular disorder and a highly structured photonic density of states. It remains unclear to what extent these additional complexities modify the usual physical picture of strong coupling developed for atoms and inorganic semiconductors. Using a microscopic quantum description of molecular ensembles in realistic multimode optical resonators, we show  that the emergence of a vacuum Rabi splitting in linear spectroscopy is a necessary but not sufficient metric of coherent admixing between light and matter. 
    In low finesse multi-mode situations we find that molecular dipoles can be partially hybridised with photonic dissipation channels associated with off-resonant cavity modes. These vacuum-induced dissipative processes ultimately limit the extent of light-matter coherence that the system can sustain.
\end{abstract}

\maketitle

\section{Introduction}

Strong coupling between a large ensemble of molecules and an optical cavity mode is still a rapidly evolving field, despite being more than 25 years old~\cite{Lidzey_PRL_1999_82_3316}. In strong coupling hybridization takes place between a molecular resonance and a cavity mode to yield two new polariton modes (states), modes that inherit characteristics of both light and matter. Two broad schemes have attracted the most attention: exciton-polaritons, where an excitonic molecular resonance is coupled to a cavity mode, and vibrational-polaritons, where a molecular vibrational mode is coupled to a cavity mode. Despite considerable progress, an underlying theoretical framework has yet to be established that provides a coherent picture of strong coupling phenomena to be given. Here we identify and explore one largely ignored ingredient, the multiple  photonic modes nature of the cavities typically employed.

One of the main attractions of molecular strong coupling is that the key phenomenon, that of an anti-crossing between a molecular resonance and a photonic mode, can be explained with a very simple model based on two coupled oscillators, one oscillator representing the molecular system (a large number of identical molecules are taken to behave as though they are a single oscillator) the other representing a single photonic (cavity) resonance. This simple picture is a powerful one, but can do little to capture a wealth of important features, including dark states, disorder, and especially material behaviour such as reactivity. It is for this reason that so much effort has been devoted to developing a wider theoretical framework. Significant progress has been made by building more realistic models of the molecular systems involved, as reviewed recently \cite{Herrera2020perspective,Sanchez2022,Mandal2023}. 

Much of the theoretical work on strong coupling in the past few years has been devoted to incorporating the complexities that arise when including more realistic numbers of molecules (typically models have $<10^3$, whilst in experiments there may be $>10^8$)~\cite{Bhuyan_AOM_2023_12_2301383}, and the presence of disorder~\cite{Khazanov_ChemSocRev_2023_4_041305}. Various approaches have been explored, examples include:
employing a Holstein-Tavis-Cummings (HTC) model \cite{Spano2015,Cwik2014,Herrera2016,Herrera_PRL_2017_118_223601} together with a Markovian~\cite{Herrera2017-PRA,delPino2018} approach for the dissipative dynamics of organic polaritons; ab-initio studies~\cite{RuggenthalerP_PRA_2014_90_012508,Flick_PNAS_2017_114_3026}; and multiscale molecular dynamics simulations~\cite{Luk_JCTC_2017_13_4324}. Whilst these theoretical approaches strive to include more realistic models for the molecular ensembles involved, little if any attention appears to have been directed towards including photonic complexities. 

It was recognised early on that the dispersion of the photon (cavity) modes was important~\cite{Agranovich_PRB_2003_67_085311}, and recent studies have focused on dispersion in connection with energy transport \cite{Ribeiro2022}. Whilst dispersion of a given photon mode is indeed important in several processes of interest such as condensation \cite{Keeling2020} and lasing \cite{Arnardottir2020}, the fact that most cavities that are currently employed in experiments support several discrete cavity modes~\cite{George_PRL_2016_117_153601}, has been largely overlooked. An unwritten assumption seems to have been that if the spacing between cavity modes (the free-spectral range) is `sufficient', then the presence of many (rather than one) photonic modes can be ignored as having minimal influence on the polariton properties. Indeed, this presumed minimal effect has led to the use of `off-resonance' modes being employed to monitor changes in the constituents within a cavity~\cite{Thomas_JCP_2024_160_204303}. Here we show explicitly that the presence of multiple photonic modes can have a significant effect on the strong coupling process. Recently the importance of this effect has been recognized in other physical implementations of cavity QED, e.g. artificial atoms in superconducting resonators~\cite{Sundaresan_PRX_2015_5_021035}. The model elucidated here is an expanded version of an outline we recently presented to  explain polariton-mediated photoluminescence in low finesse cavities~\cite{menghrajani2024}. Before looking at our multi-mode framework in detail, let us briefly mention some of the prior work on multimode cavities.

Multi-mode cavities have been employed in a large number of experiments, both for excitonic strong coupling (see for example~\cite{Coles_APL_2014_104_191108,Georgiou_JCP_2021_154_124309,Godsi_JCP_2023_159_134307}) and vibrational strong coupling (see for example~\cite{Shalabney_NatComm_2015_6_5981,Simpkins_ACSPhotonics_2015_2_1460,George_PRL_2016_117_153601,Hertzog_ChemPhotChem_2020_4_612}), but the implications of there being more than one mode have in general only involved considering the presence of single couplings, i.e. coupling between the molecular mode and each of the photonic modes. For example, when more than one photonic mode is involved then one has to take care about mode assignment due to the overlap (in energy) between different polariton bands~\cite{George_PRL_2016_117_153601}. However, as we show below, one also needs to consider how the couplings between different photonic modes alters the overall picture. One of our major findings is that such couplings can limit the extent of light-matter mixing, and may thus limit the coherence of polaritons, with possible consequences for a variety of phenomena such as photoluminescence \cite{menghrajani2024} and polariton transport~\cite{Balasubrahmaniyam_NatMat_2023_22_338}.

\begin{figure}[t]
    \centering
    \includegraphics[width=0.5\textwidth]{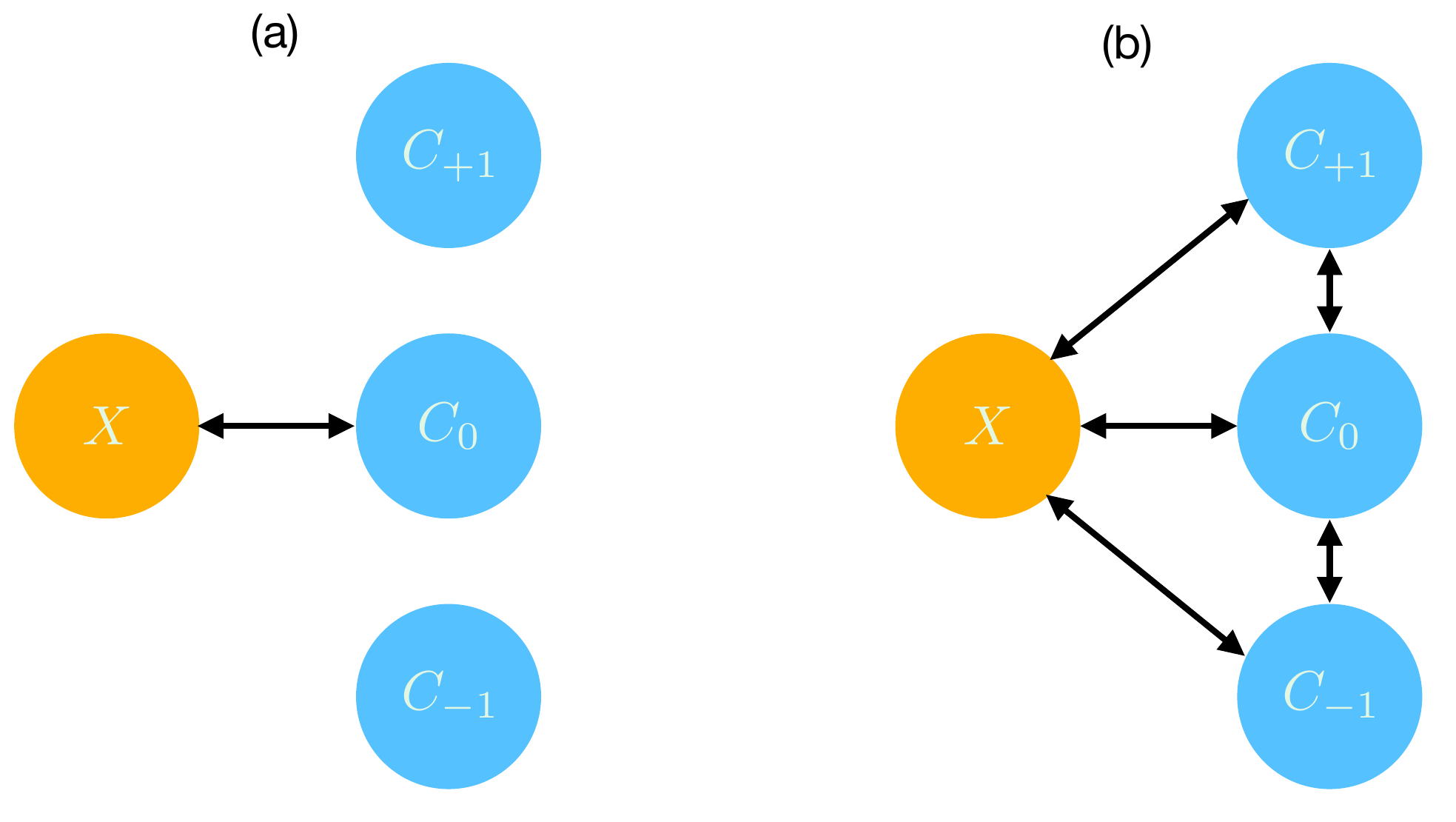}
    \caption{\textbf{Schematic:} Left, panel (a). Although several cavity modes are present ($C_{-1}$, $C_0$ and $C_{+1}$), our excitonic molecular resonance $X$ interacts with a single discrete on-resonance cavity mode, $C_0$. Right, panel (b). Now the finesse is low and there is spectral overlap between adjacent photonic (cavity) modes. The excitonic resonance interacts directly with the various cavity modes, and there is also direct interaction between adjacent photonic modes. The consequence is that the excitonic resonances is now `spread out' over more than one cavity mode.}
    \label{fig:schematic}
\end{figure}

Figure \ref{fig:schematic} provides a schematic overview of our model. In molecular strong coupling an excitonic resonance is usually considered to interact with a single cavity mode, depicted in panel (a). Although other cavity modes might be present, they do not spectrally overlap with the cavity mode being considered, they are too far detuned. However, when the finesse is low, adjacent cavity modes may overlap and may thus couple to each other and to the excitonic resonance, see panel (b). In this situation the strong coupling is no longer single mode, the exciton resonance now being `spread' over more than one cavity mode. The work reported here is the result of an investigation to explore the consequences of this `spreading'.

\section{Multi-Mode Theory of Organic Microcavities}\label{sec:Dicke-Limit}
Our aim is to build a model for molecular optical cavities that correspond to an ensemble of $N$ electronic dipole emitters coupled to the full set of resonant optical modes supported by a (planar) cavity structure. In what follows we restrict the discussion to molecular dipoles with negligible Huang-Rhys factor \cite{Hestand2018}, such that their emission properties are accurately described by considering only the vibration-less ground ($S_0\equiv g$) and first excited ($S_1\equiv e$) electronic states (see Refs. \cite{Hobson2002,Wang2019singlemolecule} for examples). The system is described by a multimode Tavis-Cummings Hamiltonian of the form ($\hbar \equiv 1$ throughout),
\begin{equation}\label{eq:Htotal}
\hat{\mathcal{H}} =  \sum_{q} \omega_q\hat a^\dagger_q\hat a_q  +\sum_{i=1}^N\omega_{i} \hat\sigma_i^{+}\hat\sigma_i^-
+\sum_{q}\sum_{i=1}^N\,g_{iq}\hat\sigma_i^{+}\,\hat a_q+g_{iq}^* \,\hat\sigma_i^{-}\hat a_q^\dagger,
\end{equation}
where $q=\{\mathbf{q}_\parallel,q_\perp\}$ is generally a composite photonic mode index describing the continuous in-plane component of the wavevector $\mathbf{q}_\parallel$ and the discrete transverse component $q_\perp$ of confined electromagnetic modes in planar cavities \cite{Skolnick1998}. The local dipole transition operators between ground $\ket{g_i}$ and excited state $\ket{e_i}$ are defined as $\hat \sigma_i^-=\ket{g_i}\bra{e_i}$ for emission and $\hat \sigma_i^+= (\hat \sigma_i^-)^\dagger$ for absorption. The electronic transition frequency $\omega_i$ is in general inhomogeneously distributed, although  there are examples of organic emitters with negligible inhomogeneous broadening \cite{Wang2019}. Bosonic cavity field operators are $\hat a_q$, and the local mode-dependent Rabi couplings are denoted by $g_{iq}$. 

Ideal planar cavities of length $L$ have photon dispersion $\omega_q=(c/n_{\rm d})({q_\parallel^2+q_\perp^2(m)})^{1/2}$, where $q_{\perp}(m)=m\pi/2L$,  with $m\geq 1$ an integer, which determines the discrete set of allowed cavity mode energies at normal incidence ($q_\parallel =0$); $c$ is the speed of light and $n_d$ the (real) dielectric constant of the intracavity medium. The  mode dispersion with respect to the in-plane wavevector $\mathbf{q}_\parallel$ determines the propagation properties of the normal modes of the coupled system, which for strong  light-matter coupling correspond to exciton-polaritons \cite{Lidzey1998,Barnes1998,kena-cohen2008}, and the free spectral range (FSR) between adjacent modes $\Delta_{q} \equiv \omega_{q+1}-\omega_q$, is controlled at normal incidence ($q_\parallel=0$) by the cavity length as $\Delta = c\pi/2n_dL$. Throughout this work, we neglect dispersion and only study system properties at $q_\parallel =0$. 

The explicit multimode structure of Eq. (\ref{eq:Htotal}) generalizes early approaches that simplify the mode structure of the microcavity to having a single dispersionless mode $\hat a$ \cite{Herrera2017-review}, or a dispersive mode $\hat a_{q_\parallel}$ in a cavity with \emph{infinite} FSR. Such simplifications are often introduced as a necessity in favour of capturing relevant aspects of complexity of the internal degrees of freedom of the molecular dipole emitters, such as high-frequency vibrations \cite{Cwik2014,Spano2015,Herrera2016,Herrera2017-PRA,Herrera_PRL_2017_118_223601,Spano2020}, electron tunnelling \cite{Hagenmuller2017}, or the role of static disorder in establishing quantum transport regimes \cite{Engelhardt2023,Chavez2021,Ribeiro2022}. Single-mode or single-branch theories are intrinsically limited with respect to their ability to describe the influence of multiple transverse photonic modes in realistic organic microcavites with finite FSR, $\Delta$, as the value of $\Delta$ is not much larger than the frequency separation between lower and upper polaritons, i.e. the Rabi splitting $\Omega_R$. For some systems $\Delta$ is even smaller than $\Omega_R$~\cite{Kena-Cohen2013}. 

\section{Microcavities as Open Quantum Systems}

We model the organic microcavity microscopically as an open quantum system described by a Lindblad quantum master equation for the light-matter reduced density matrix $\hat \rho$, given by~\cite{Breuer-book},
\begin{eqnarray}\label{eq:qme}
\frac{d}{dt}\hat \rho &=& -i[\hat{\mathcal{H}},\hat \rho]+\sum_q \frac{\kappa_q}{2}\left(2\hat a_q \hat \rho\hat a_q^\dagger- \hat a_q^\dagger \hat a_q \hat \rho - \hat \rho  \hat a_q^\dagger \hat a_q\right) \nonumber\\
&&+ \sum_i \frac{\gamma_i}{2}\left(2\hat \sigma^-_i \hat \rho\hat \sigma^+_i- \hat \sigma_i^+ \hat \sigma^-_i \hat \rho - \hat \rho \hat \sigma_i^+ \hat \sigma_i^-\right),
\end{eqnarray}
where $\kappa_q$ is the bare radiative decay rate of the $q$-th cavity mode and $\gamma_i$ the bare spontaneous decay rate of the $i$-th electronic dipole excitation, which includes radiative and non-radiative contributions. For simplicity, we ignored cross-terms that could dissipatively couple different cavity modes or different molecules, under the assumption that  direct diagonal relaxation channels are much faster. 

The Lindblad master equation is the basis for deriving effective non-unitary  propagators that unravel the state evolution as an ensemble of  wavefunction trajectories \cite{Molmer:93}. For the light-matter state ansatz,
\begin{equation}
\ket{\Psi(t)}\approx \sqrt{1-\xi(t)}\ket{\psi^{(0)}(t)} + \sqrt{\xi(t)}\ket{\psi^{(1)}(t)},    
\end{equation}
with $\xi\ll 1$, as is relevant for weakly excited microcavities, we can ignore stochastic quantum jumps coming from the recycling terms of the Lindblad equation \cite{Herrera2017-PRA}, and account for dissipation as an exponential decay of the excited state population. This is equivalent to rewriting Eq. (\ref{eq:qme}) as, 
\begin{equation}
\frac{d}{dt}\hat \rho = -i[\hat{\mathcal{H}}_{\rm eff},\hat \rho] + \mathcal{L}_1[\hat \rho],
\end{equation}
with an effective non-Hermitian Hamiltonian, 
\begin{equation}\label{eq:Heff}
\hat{\mathcal{H}}_{\rm eff} = \hat{\mathcal{H}} -\frac{i}{2}\sum_{q}\kappa_q\hat a^\dagger_q\hat a_q -\frac{i}{2}\sum_i\gamma_i \hat\sigma_i^{+}\hat\sigma_i^-.
\end{equation}
As mentioned above, we ignore the recycling terms $\mathcal{L}_1[\hat \rho]\equiv \sum_q \kappa_q \hat a_q \hat \rho\hat a_q^\dagger+\sum_i \gamma_i \hat \sigma^-_i \hat \rho\hat \sigma^+_i$. This simplified approach to the open system dynamics effectively reduces the problem to one of solving a non-Hermitian Schrodinger equation with the effective Hamiltonian $\hat{\mathcal{H}}_{\rm eff}$.

Since the ground state $\ket{\psi^{(0)}}$ has no electronic or photonic excitations, the dynamics of polaritons is fully determined by the excited electron-photon wavefunction 
\begin{equation}\label{eq:ansatz}
\ket{\psi^{(1)}} = \sum_i c_i^{(0)}\ket{e_i}\ket{\{0_q\}}+\sum_q\,c_{q}^{(1)}\ket{g_1g_2,\ldots,g_N}\ket{1_{q}},
\end{equation}
where $\ket{\{0_q\}}$ is the multi-mode cavity vacuum, $\ket{e_i}$ describes a single excitation in the $i$-th molecule with all other dipoles in the ground state. $\ket{1_q}$ describes a single photon in the $q$-th transverse mode, all other modes being empty.

\section{Strong Coupling in High-Finesse Cavities}\label{sec:high finesse}

Consider $N$ molecular dipoles coupled near resonantly with a $q=0$ mode of frequency $\omega_0$ and decay rate $\kappa_0$. Higher and lower order cavity modes are detuned from the central frequency by $\Delta_q=\omega_q-\omega_0$, with $\Delta_q>0$ for higher-order and $\Delta_q<0$ for lower-order modes. They also have bandwidths that in general differ from  $q=0$ by $\Delta \kappa_q = \kappa_q-\kappa_0$. In the large finesse limit, $|\Delta_q|\gg\Omega_0$, with $\Omega_0=\sqrt{N}g_0$ being the single-mode Rabi coupling strength for a homogeneous molecular ensemble, dipole excitations cannot exchange energy effectively with higher and lower order cavity modes. (Note, the single mode Rabi splitting $\Omega^{SM}_R$ is related to the single mode Rabi coupling $\Omega_0$ as $\Omega^{SM}_R$ = 2$\Omega_0$.) Consequently, light-matter hybridization leading to polariton formation only occurs in the vicinity of the near-resonant $q=0$ mode, see panel (a) of figure \ref{fig:schematic}. However, far de-tuned modes do have an effect via second order (two photon) processes, something we look at next.

Far-detuned higher and lower-order modes evolve on a timescale of $1/\Delta_q$, which is much faster than the Rabi oscillation period $\tau_R\sim 1/\Omega_0$ between the near-resonant mode and molecular excitations. These fast-oscillating mode variables thus adiabatically adjust to  the dynamics of the near-resonant manifold, which affects the process of polariton formation around $q=0$. This can be understood as the emergence of Raman-type processes in which molecules absorb and re-emit virtual photons from higher and lower-order modes. 
These two-photon processes result in a change to the energetics; a single-molecule frequency shift of the form, 
\begin{equation}\label{eq:ground shift}
\Gamma_j^{''} = -\sum_{q\neq 0} |g_{jq}|^2\frac{\Delta_q}{\Delta_q^2+(\Delta \kappa_q/2)^2},
\end{equation}
and an effective inter-molecular coupling with interaction energy given by, 
\begin{equation}\label{eq:effective dipole-dipole}
J_{ij}'' = -\sum_{q\neq 0} g_{iq}^*g_{jq} \frac{\Delta_q/2}{\Delta_q^2+(\Delta \kappa_q/2)^2}.
\end{equation}
The sign of the contributions per mode in these expressions is different for lower-order modes (blue shift and repulsive interaction) and higher-order modes (red shift and attractive interactions). In general, $J_{ij}''$ is a complex-valued quantity depending on the relative phase of the Rabi frequency at the location of the two dipoles.

Cavity-induced frequency shifts and inter-dipole interactions induced by far-detuned modes are well-known from atomic physics \cite{Raimond2001} and have been used for quantum state preparation in high-quality resonators \cite{Leroux2010}. Organic microcavities are qualitatively different from atomic cavities in that their quality factors are much lower, typically $Q\sim 1-10$, and changes in bandwidth $\kappa_q$ with mode order can be large. This is particularly true for modes close to the region where absorption of metal mirrors cannot be neglected \cite{Thomas_AdvMat_2024_36_2309393}. 
Therefore, in addition to the changes in frequency and interaction energy discussed above, the dispersive interaction of molecular dipoles with far-detuned lossy modes also changes the single-molecule dipole decay rates by, 

\begin{equation}\label{eq:effective one-body decay}
\Gamma_j' = -\sum_{q\neq 0} |g_{jq}|^2\frac{ \Delta\kappa_q/2}{\Delta_q^2+( \Delta\kappa_q/2)^2}, 
\end{equation}
and establishes the two-body loss rate,
\begin{equation}\label{eq:two-body loss}
J_{ij}' =-\sum_{q\neq 0} g_{iq}^*g_{jq} \frac{ \Delta\kappa_q/2}{\Delta_q^2+( \Delta\kappa_q/2)^2}.  
\end{equation}
These are again signed quantities summed over all available modes, whose contribution depends on the relative bandwidths $\Delta \kappa_q$. If all relevant modes have bandwidths $\kappa_q$  equal to the resonant ($q=0$) mode, then $\Delta \kappa_q\approx0$ and no second-order corrections to the decay rates are expected. The bandwidth mismatch of sub-wavelength cavities thus introduces a phenomenology that is not present in other cavity QED systems, we discuss one key aspect next.  

The dispersive relaxation channels discussed above involving material degrees of freedom could, in principle, alter the ability to establish strong coupling with the central $q=0$ mode. To assess this, consider a simplified homogeneous scenario in which all dipoles are identical (Dicke regime) and the polariton wavefunction in Eq. (\ref{eq:ansatz}) reduces to,
\begin{equation}\label{eq:polariton state SM}
\ket{\psi^{(1)}}= \beta\ket{X}\ket{0}+\alpha\ket{g_1g_2,\ldots,g_N}\ket{1},
\end{equation}
where $\ket{X}=\sum_j\ket{e_j}/\sqrt{N}$ is the fully-symmetric excitonic state and the  Fock states $\{\ket{0},\ket{1}\}$ refer to the central $q=0$ mode. The dynamics of the state vector $\mathbf{x}=[\alpha,\beta]^{\rm T}$ can be written as $\dot{\mathbf x}=-iM\mathbf{x}$, where, 
\begin{equation}\label{eq:adiabatic Ham}
M 
= \left( \begin{array}{cc}
0 & \Omega_0\\
\Omega_0& -\delta_N -i \Delta\Gamma_N
\end{array}\right),
\end{equation}
is the dynamical matrix whose complex eigenvalues $\lambda=E+i\Gamma/2$ give the polariton energies, $E$, and bandwidths, $\Gamma$.  $\Omega_0=\sqrt{N}g_0$ is the Rabi coupling strength.  The effective detuning $\delta_N$ and bandwidth mismatch $\Delta\Gamma_N$ can be written as
\begin{eqnarray}
\delta_N&=& \delta_0-\Gamma{''}-NJ{''}\label{eq:delta N}\\
\Delta\Gamma_N &=&  -\Delta\gamma/2 +\Gamma' + NJ{'} \label{eq:Gamma N},
\end{eqnarray}
where $\delta_0=\omega_0-\omega_e$ and $\Delta \gamma=\kappa_0-\gamma$ are the bare detuning and bandwidth mismatch between the $q=0$ mode and the dipole resonance. The one-body and two-body energy shifts $\Gamma''$ and $NJ''$ contribute to the detuning of the $q=0$ mode from the molecular resonance. The one- and two-body decay rates $\Gamma'$ and $NJ'$ contribute to the bandwidth mismatch. The microscopic derivation of Eq. (\ref{eq:adiabatic Ham}) starting from Eq. (\ref{eq:Heff}) is given in the appendix.

The real part of the eigenvalues of $M$ give the lower and upper polariton frequencies, $E_{\rm LP}$ and $E_{\rm UP}$, respectively. The Rabi splitting $\Omega_R\equiv E_{\rm UP}-E_{\rm LP}$ can thus be written as
\begin{equation}\label{eq:Rabi splitting}
    \Omega_R= {\rm Re}\sqrt{(NJ''-\delta_0-i(NJ'-\Delta\gamma/2))^2+4\Omega_0^2},
\end{equation}
where $\Gamma'$ and $\Gamma''$ are neglected in the thermodynamic limit~\footnote{The thermodynamic limit being $N\rightarrow\infty$, with $\Omega_0=\sqrt{N}g_0$, $NJ'$ and $NJ''$ finite.}. Equation (\ref{eq:Rabi splitting}) gives the usual strong coupling result $\Omega_R=2\Omega_0$ for infinite finesse, $\Delta_q\rightarrow\infty$ and finite $N$, since $NJ'\sim 1/\Delta_q^2$ and $NJ''\sim 1/\Delta_q$.

\begin{figure*}[t]
    \centering
    \includegraphics[width=\textwidth]{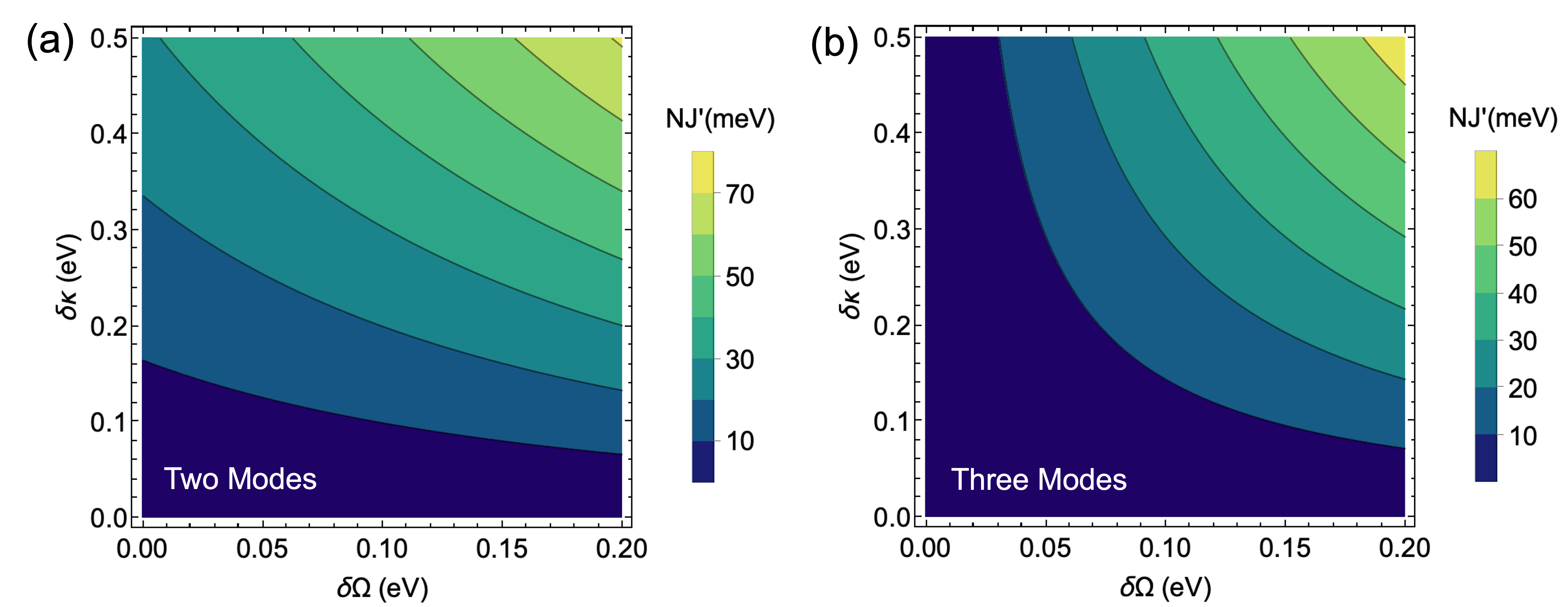}
    \caption{\textbf{Modified polariton decay in high finesse cavities:} Two-body contribution to the polariton decay rate $NJ'$, as a function of the change in Rabi coupling $\delta\Omega$ and change in bandwidth $\delta\kappa$ relative to a reference $q=0$ strongly coupled resonant mode for two cases; (a) two-mode cavity where $q=0$ is the lowest order mode; (b) three-mode cavity with one lower and one higher-order mode. We use $\Omega_0=0.35$ eV and  $\Delta= 1.0$ eV.}
    \label{fig:two-body loss}
\end{figure*}

To estimate the magnitude of $NJ'$ for typical high-finesse cavities ($|\Delta_q|>\Omega_R$), consider a model three-mode cavity with a central $q=0$ mode at $\omega_0$, a lower-order mode ($q=-1$) detuned from $q=0$ by $\Delta_{-1}=-\Delta$ and a higher-order mode ($q=+1$) detuned by $\Delta_{+1}=\Delta$, with $\Delta>0$. The mode-dependent decay rates are $\{\kappa_-, \kappa_0, \kappa_+\}$, respectively. Without losing generality, we assume linear scaling of the bandwidth with mode order, i.e., $\kappa_q = \kappa_0 +q \,\zeta$ with positive $\zeta$ for increasing bandwidth and negative otherwise. We also allow the Rabi coupling strength to depend on mode order as $\Omega_{\pm 1}\equiv \sqrt{N}g_{\pm 1}$, for collective coupling of dipoles to $q=\pm1$ cavity modes. From Eq. (\ref{eq:two-body loss}), the two-body rate can be written as, 
\begin{equation}\label{eq:J decay homo}
NJ{'} = \left(\Omega_{-1}^2-\Omega_{+1}^2\right)\frac{\zeta/2}{{\Delta}^2+\zeta^2/4}\approx \frac{\zeta f\,\Omega_0^2}{{\Delta}^2+\zeta^2/4},
\end{equation}
where in the second equality we used $\Omega_{\pm 1}=\Omega_0(1\pm f)$ with $|f|<1$. This contribution to the polariton decay can either increase or decrease the bandwidth of LP and UP resonances around the $q=0$ mode, depending on the sign of $\zeta f$.

Figure \ref{fig:two-body loss} shows the magnitude of $NJ'$ from Eq. (\ref{eq:J decay homo}) as a function of the variation in Rabi coupling per mode $\delta\Omega=f\Omega_0$ and the variation in bandwidth $\delta\kappa=\zeta$, estimated for a system with $\Omega_0=0.35$ eV and FSR $\Delta= 1.0$ eV. We consider a two-mode cavity (panel a), where $q=0$ is the lowest-order mode ($\Omega_{-1}=0$) and modes beyond $q=1$ are ignored, as well as the three-mode scenario (panel b). In general, the magnitude of $NJ'$ is smaller than $k_{\rm B}T\approx 26\,{\rm meV}$ for multi-mode microcavities with relatively weak mode-dependence of the Rabi coupling and photon bandwidth  ($\delta\Omega\sim \delta\kappa \sim 10^{-2}$ eV \cite{menghrajani2024})), but the analysis above is general and larger polariton bandwitdh modifications could be expected for other high finesse photonic structures with greater cavity bandwidths and for coupling variations with mode order.

In summary, the presence of far off-resonance cavity modes can introduce adiabatic corrections to the Rabi splitting $\Omega_R$ established in strong coupling. Such corrections originate from coherent and incoherent two-photon Raman-type processes in which dipoles scatter virtual photons from far-detuned higher and lower-order modes, primarily leading to changes in the dipole bandwidth (Fig. \ref{fig:two-body loss}). Since adiabatic corrections to $\Omega_R$ scale as $(\Omega_0/\Delta)^2$, it can be difficult to measure their contribution in typical high finesse Fabry-Perot microcavities ($\mathcal{F}\equiv \Delta/\kappa_0\gtrsim 10$, $\Omega_0/\Delta\lesssim 0.1$).

\section{Strong Coupling in Low-Finesse Cavities}\label{sec:low finesse}

The adiabatic elimination procedure in the previous Section is strictly valid for $\Omega_R/\Delta\ll 1$ and breaks down if the Rabi splitting is not much smaller than the FSR, even when the bare cavity finesse is nominally high ($\mathcal{F}\ge10$). 

For microcavities with lower finesse, there is no significant separation of scales between $\kappa$, $\gamma$, $\Omega_0$ and $\Delta$, although typically $\kappa\sim \gamma<\Omega_0<\Delta$ in strong coupling \cite{menghrajani2024}. For the light-matter system discussed above, with $N$ identical dipoles at $\omega_e$ resonant with a reference $q=0$ cavity mode at $\omega_0$, having Rabi coupling strength $\Omega_0$, the frequency separations $\Delta_{\pm 1}$ of adjacent higher-order and lower-order modes ($q\pm 1$) are comparable with the bare Rabi couplings $\Omega_{\pm 1}$ to those modes and thus the direct coupling of  dipoles to neighbouring modes needs to be considered.

For the homogeneous dipole ensemble, the simplest extension of the light-matter state vector includes the next-order modes $q=\pm 1$, i.e., $\mathbf{x}=[\alpha_{-1},\alpha_0,\alpha_{+1},\beta]^{\rm T}$, where $\alpha_q$ is the photon amplitude in the $q$-th mode. The dynamical matrix for this (3+1) system can be written as,
\begin{equation}\label{eq:H three-mode}
M 
= \left( \begin{array}{cccc}
d_{-1} &0&0& e_{-1}\\
0 &0&0& e_{0}\\
0 &0& d_{+1}& e_{+1}\\
e_{-1} &e_{0}&e_{+1}& p\\
\end{array}\right),
\end{equation}
where we used the simplified notation $d_{q}=\Delta_q-i\Delta\kappa_q/2$, $e_q=\Omega_q$ and $p=-\delta+i\Delta\gamma/2$. From the eigenvalues of $M$, $\lambda=E+i\Gamma/2$, the coupled energies $E$ and decay rates $\Gamma$ are obtained. In general, $\lambda$ is a  root of the polynomial,
\begin{equation}\label{eq:phi}
\Phi(\lambda)\equiv p-\lambda +\frac{e_0^2}{\lambda}- \frac{e_{-1}^2}{d_{-1}- \lambda} -\frac{e_{+1}^2}{d_{+1}- \lambda},
\end{equation}
which for $\Omega_{\pm1}=0$ gives the quadratic eigenvalue equation, 
\begin{equation}\label{eq:quadratic sc}
\lambda(\delta-i\Delta\gamma/2-\lambda)-\Omega_0^2=0 ,      
\end{equation}
that is often used to derive conditions for strong coupling in the single-mode picture. In particular, for  resonant bandwidth-matched light-matter interaction with the $q=0$ mode ($\delta=0$, $\gamma=\kappa$), Eq. (\ref{eq:quadratic sc}) gives the LP and UP frequencies $E_\pm=\pm \Omega_0$ ($\Omega_R=2\Omega_0$) with decay rates $\Gamma_\pm =\kappa$ \footnote{In the frame where $M$ in Eq. (\ref{eq:H three-mode}) is defined, coupled bandwidths are given by $\Gamma=\kappa+2{\rm Im}[\lambda]$, where $\lambda$ is an eigenvalue of $M$, see the appendix.}.

\begin{figure*}[t]
    \centering
    \includegraphics[width=\textwidth]{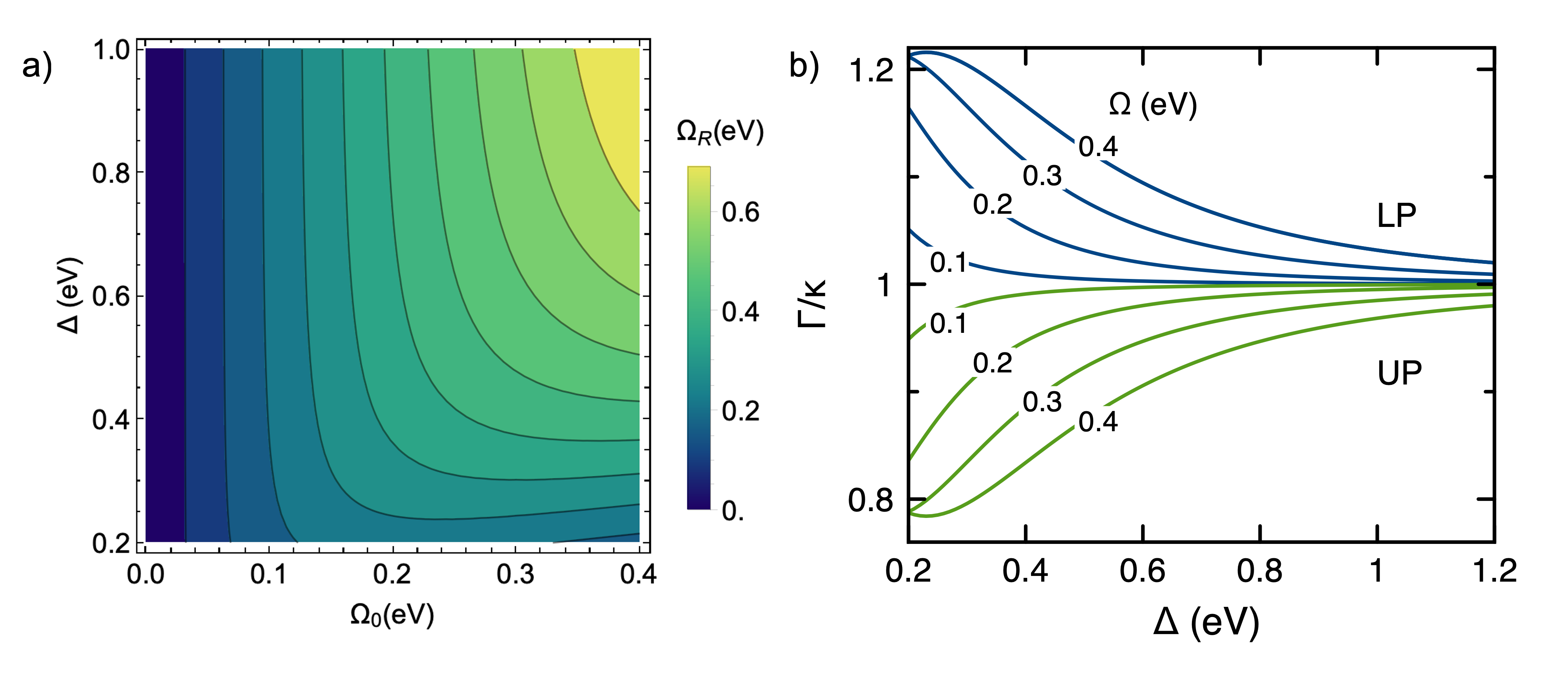}
    \caption{\textbf{Modified Rabi splitting and bandwidth in low finesse cavities:} (a) Rabi splitting $\Omega_R$ as function of the bare Rabi coupling $\Omega_0$ and free-spectral range $\Delta$ for a three-mode cavity ($q=\{0,\pm1\}$) with the central mode on exact resonant with the dipole transition; (b) Decay rates $\Gamma$ of the Lower polariton (LP, blue lines) and upper polariton (UP, green lines) in units of $\kappa$, as functions of $\Delta$ for different values of the bare Rabi couping $\Omega_0$, for a linear variation of the mode bandwidths $\kappa_q=\kappa_0+q\zeta$.  We use $\delta=0$, $\gamma=\kappa=0.1\,{\rm eV}$ and $\zeta=-0.1$ eV.  -mode cavity with one lower and one higher-order mode. We use $\Omega_0=\Omega_{\pm 1}=0.35$ eV, $\kappa=\gamma=0.15 \,{\rm eV}$, $\zeta=-0.1\, {\rm eV}$.}
    \label{fig:splitting and bandwidth}
\end{figure*}

Direct coupling of dipoles to the $q=\pm 1$ modes modifies the LP and UP energies and bandwidths. In the appendix we derive a general expression for the lowest-order corrections to the energies and bandwidths of the single-mode polariton problem, due to the presence of neighbouring cavity modes. These corrections scale nonlinearly with the Rabi couplings $\Omega_{\pm1}$ and mode detunings $\Delta_{\pm 1}$, and have a strong dependence with the change in bandwidth $\Delta\kappa_q$ between different cavity modes. For a resonant, bandwidth-matched interaction with the $q=0$ mode, and ignoring possible changes of the Rabi coupling for different cavity modes ($\Omega_q=\Omega$), the modified Rabi splitting and polariton bandwidths around the $q=0$ mode can be approximated as,
\begin{equation}\label{eq:polariton splitting}
\Omega_R =2\Omega\left[1-\frac{\Omega^2(\Omega^2+\Delta^2)}{(\Omega^2+\Delta^2)^2+\Delta^2\zeta^2}\right],
\end{equation}
and,
\begin{equation}\label{eq:polariton bandwidths}
 \Gamma_\pm=\kappa \pm 2\Omega \frac{\Omega^2\Delta\zeta}{(\Omega^2+\Delta^2)^2+\Delta^2\zeta^2}, 
\end{equation}
where again $\Delta_{\pm1}=\pm\Delta$ and $\kappa_q=\kappa_0+q\zeta$ are assumed. These expressions reduce to the single-mode case, when $\Omega\ll\Delta$ and $|\zeta|\ll \Delta$, which is the high-finesse regime discussed in the previous section.  Although the correction to the splitting also scales as $(\Omega/\Delta)^2$ for  $\Delta\gg \Omega$ as in the high-finesse problem, now the presence of nearby modes in finite finesse cavities {\it directly} modifies the polariton energies $E_\pm$ around the dipole resonance via level repulsion. In contrast, adiabatic corrections introduce an overall dipole shift via two-photon Raman processes which can in principle be compensated for by tuning the cavity frequency.

\begin{figure*}[t]
    \centering
    \includegraphics[width=\textwidth]{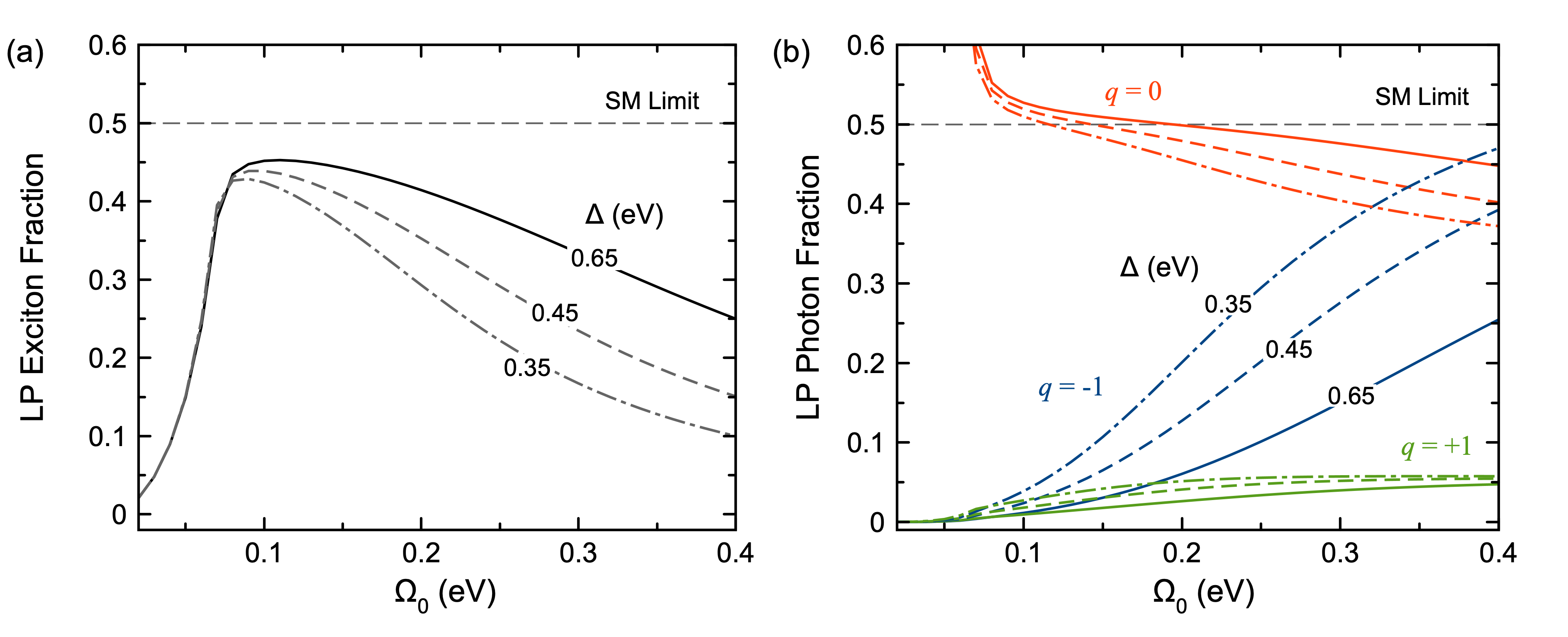}
    \caption{\textbf{Modified polariton wavefunctions in low finesse cavities:} (a) \textbf{Exciton fraction} of the LP state as a function of the bare Rabi coupling for a three-mode cavity with realistic parameters ($\Delta \approx 0.65$ eV \cite{Thomas_AdvMat_2024_36_2309393}). The single-mode (SM) limit for $q=0$ resonant with the dipoles is marked. Results for other values of $\Delta$ are also shown, keeping other parameters the same. (b) \textbf{Photon fraction} per $q$-mode of the LP state as a function bare Rabi coupling for different values of $\Delta$, using the same parameters in (a).}
    \label{fig:multi-mode state contents}
\end{figure*}

Similar to the adiabatic corrections in Eq. (\ref{eq:J decay homo}), the changes of the polariton decay rates $\Gamma_\pm$  predicted for low-finesse cavities also scale linearly with the difference in bandwidth $\Delta\kappa_q$ of nearby modes relative to the near-resonant $q=0$ mode. However, whilst adiabatic bandwidth corrections vanish for systems with mode-independent Rabi couplings, Eq. (\ref{eq:polariton bandwidths}) suggests that the mode-order dependence of the bare bandwidths is more important than variations in the field profile (Rabi coupling) of different cavity modes to establish the bandwidths of the LP and UP resonances.

Figure \ref{fig:splitting and bandwidth}a shows the Rabi splitting $\Omega_R$ (in eV) predicted by Eq. (\ref{eq:polariton splitting}) as function of the bare Rabi coupling $\Omega_0$ with the central mode ($\Omega_q=\Omega$) and the FSR $\Delta$, for a system with decreasing mode bandwidth with increasing mode order ($\zeta=-0.1$ eV). Even for relatively large values of $\Delta\sim 1 {\rm eV}$, the direct interaction of dipoles with $q=\pm 1$ modes significantly reduces the multi-mode Rabi splitting from the usual single-mode value $\Omega_{R}^{\rm SM}= 2\Omega_0$. In practical terms, the requirements for establishing a light-matter interaction strength that gives a desired polariton splitting become more demanding as the finesse decreases. Figure \ref{fig:splitting and bandwidth}b shows the complementary effect on the LP and UP bandwidths $\Gamma_{\pm}$ given by Eq. (\ref{eq:polariton bandwidths}), as functions of $\Delta$. The bare mode bandwidth is $\kappa=0.15$ eV and a small linear decrease of the bandwidth with mode order is assumed ($\zeta=-50$ meV). The polariton level (UP) closer in frequency to the narrower mode ($q=+1$), becomes narrower as $\Delta$ decreases, and the level (LP) closer to the broader mode ($q=-1$) broadens. Even for relatively large mode separations ($\Delta= 1.2 $eV, $\mathcal{F}\approx 8$), the LP and UP bandwidths are asymmetric and can differ significantly from the single-mode prediction  $\Gamma_\pm^{\rm SM} =(\kappa+\gamma)/2$. The deviation from $\Gamma_{\pm }^{\rm SM}$ grows with increasing coupling strength $\Omega_0$. 

In addition to the changes in Rabi splitting and polariton bandwidths introdcude by the direct interaction of dipoles with neighbouring modes, the microscopic composition of the LP and UP polariton wavefunctions can be very different in low-finesse cavities relative to a single-mode picture. Understanding this can improve the ability to control the emission properties of organic microcavities \cite{Herrera2017-PRA,Herrera_PRL_2017_118_223601}.

In an ideal single-mode cavity under strong coupling conditions $\Omega_R>(\kappa_0+\gamma)/2$, the LP and UP states can be accurately described Eq. (\ref{eq:polariton state SM}) with $|\beta|=|\alpha|=1/2$, i.e., equal exciton and photon content. For the three-mode problem centered around $q=0$, molecular dipoles can also admix with the lower- and higher-order modes $q=\pm 1$. Therefore, a more realistic description of LP and UP states would be
\begin{eqnarray}\label{eq:three-mode state}
\ket{\psi} &=& \beta\ket{X}\ket{0_{0}}\\
&&+\ket{G}\otimes \left(\alpha_{-1} \ket{1_{-1}}+\alpha_0\ket{1_{0}}+\alpha_{+1}\ket{1_{+1}}\right)\nonumber
\end{eqnarray}
where $\ket{n_q}$ denotes a Fock state of the $q$-th cavity mode and $\ket{G}=\ket{g_1g_2,\ldots,g_N}$.
 $|\beta|^2$ is the exciton fraction and $|\alpha_q|^2$ the fraction of the $q$-th cavity mode in the polariton state. Equation (\ref{eq:three-mode state}) highlights that dipoles exchange energy and coherence with a {\it single-photon wavepacket}, not an individual Fock state. Since $\ket{\psi}$ is normalized, the photon fraction associated with the near-resonant mode $|\alpha_0|^2=1-|\beta|^2-|\alpha_{-1}|^2-|\alpha_{+1}|^2$ is always smaller than  the single-mode limit ($\Delta\rightarrow\infty, \alpha_{\pm 1}\rightarrow0$).  The continuous-variable version of the single-photon wavepacket in Eq. (\ref{eq:three-mode state}) arises naturally when describing dipole emission in macroscopic QED \cite{Feist2021,Medina2021}.

Figure \ref{fig:multi-mode state contents}a shows the material (exciton) content of the LP state as a function of the bare coupling $\Omega_0$, obtained from the eigenvectors of the three-mode matrix $M$ in Eq. (\ref{eq:H three-mode}), parametrised with realistic frequencies and bandwidths from Ref. \cite{Thomas_AdvMat_2024_36_2309393} ($\omega_e=2.15$ eV, $\gamma=0.37$ eV, $\omega_{-1}=1.45$ eV, $\omega_0=2.14$ eV, $\omega_{+1}=2.76$ eV, $\kappa_{-1}=38$ meV, $\kappa_{0}=90$ meV, $\kappa_{+1}=90$ meV). $\Omega_0=\Omega_{\pm 1}$ is a free parameter. The $q=-1$ mode is significantly narrower than $q=0$, but no significant variation is seen for $q=1$. The standard single-mode picture of  strong coupling suggests that for Rabi splittings $\Omega_{R}>(\kappa_0+\gamma)/2\approx 0.24$ eV ($\Omega_0>0.12$ eV) the exciton fraction should be $|\beta|^2\approx 0.5$. In contrast, the exciton fraction of the LP state in Fig. \ref{fig:multi-mode state contents}a  {\it decreases} with increasing Rabi coupling. We also show results for a hypothetical scenario where the lower and higher-order mode frequencies are varied as $\omega_{\pm 1}'=\omega_{\pm 1}\mp \varepsilon$ (reducing $\Delta$), with all other parameters kept constant \footnote{In real microcavity structures, the mode bandwidths and Rabi couplings also vary with mode frequency.}. We find that even for moderate values of cavity finesse $\Delta/\kappa\sim 4-5$, the exciton fraction of the LP can be lower than 30\% even when the Rabi coupling $\Omega_0$ exceeds the bare bandwidths $\kappa_0$ and $\gamma$ ($\Omega_0>0.3$ eV, $\Omega_R> 0.5$ eV). 

Figure \ref{fig:multi-mode state contents}b shows a complementary view of the photon content per mode of the LP state, for the same system parameters in Fig. \ref{fig:multi-mode state contents}a. As the bare Rabi coupling increases, the contribution of the $q=0$ mode decreases below the single-mode limit and the $q=-1$ contribution increases significantly. The changes in the $q=0$ and $q=-1$ components with $\Omega_0$ are stronger with decreasing inter-mode separation $\Delta$. The higher-order $q=+1$ state component is less sensitive to $\Omega_0$ and $\Delta$, because it is further detuned from the LP.

\section{Conclusion and Outlook}\label{sec:Conc-Outlook}

It has been understood for many years that taking account of some of the real world complexities, especially disorder, is important in building up a full conceptual model of molecular strong coupling. In this contribution we have shown that it is also important to take into account off-resonance photonic modes supported by a cavity if one is to properly understand molecular strong coupling. In particular we have shown that in low-finesse situations the extent of light-matter mixing (hybridisation) is altered. When adjacent photonic modes are spectrally overlapped then the molecular content (here we considered excitons) is spread over several photonic modes, resulting in a lower matter content in any given polariton mode. This conclusion is supported by a similar finding in the context of circuit QED~\cite{Sundaresan_PRX_2015_5_021035}. We have also shown that the extent of the Rabi-splitting can be curtailed by the effect of extra photonic modes.

Looking ahead it will be important to explore these issues further, especially in conjunction with experiment. We have already made a start in this direction~\cite{menghrajani2024}, where we monitored luminescence from a range of planar samples, since luminescence probes the extent of light-matter mixing more directly that, for example, reflectivity. It would be useful to build on this start with a more systematic study. For example, one could envisage a series of experiments on Fabry Perot planar cavities that employ metal mirrors, where the metal-mirror thickness is altered to control the cavity mode-width, and hence the finesse. 

Regarding development of the model that we have outlined here, future extensions should include frequency disorder of dipole transitions, which in most cases is the dominant contribution to material absorption lineshape \cite{Spano2010}. Disorder leads to the formation of semi-localized states in the spectral region where the uncoupled dipoles also absorb \cite{Dubail2022,Wellnitz2022}, adding complexity to the analysis of spectroscopic signals not present in the homogeneous dipole (Dicke) models we discussed here. We anticipate that disorder will introduce quantitative changes to the dependence with finesse of the polariton splitting and polariton bandwidths relative to the homogeneous model predictions here, but we the qualitative physics should remain. We already find evidence of the reduction of the exciton content of exciton-polariton states due to multimode light-matter interaction in a realistically disordered system \cite{menghrajani2024}. A more general theoretical framework should be able to treat the simultaneous coupling of molecular dipoles to multiple discrete transverse modes and continuous in-plane momenta in a microcavity. Such a theory would be complex, but may enable studies of controlled excitation transport along a polariton branch by possibly driving off-resonant coupled photonic branches with external fields.   
\\
\vspace{4cm}

\section{Acknowledgements}
WAB acknowledge the support of European Research Council through the photmat project (ERC-2016-AdG-742222 :www.photmat.eu). FH acknowledges the support of the Royal Society through the award of a Royal Society Wolfson Visiting Fellowship, and ANID through grants Fondecyt Regular 1221420, Millennium Science Initiative Program ICN17\_012, and the Air Force Office of Scientific Research under award number FA9550-22-1-0245. 

\begin{widetext}

\appendix
\section{Adiabatic elimination of far-detuned cavity modes}
\label{app:adiabatic elimination}

In this section we derive Eq. (\ref{eq:adiabatic Ham}) in the main text. \\

The non-degenerate ground state has no excitations in the electronic and cavity modes and is given by $\ket{\psi^{(0)}}=\ket{g_1g_2,\ldots,g_N}\ket{0_{-K},\ldots,0_{-1},0_0,0_{+1},\ldots,0_K}$. $M=2K+1$ discrete transverse modes are considered. The single-excitation electron-photon wavefunction can be written as 
\begin{equation}\label{eq:ansatz SM}
\ket{\psi^{(1)}} = \sum_i c_i^{(0)}\ket{e_i}\ket{\{0_q\}}+\sum_q\,c_{q}^{(1)}\ket{g_1g_2,\ldots,g_N}\ket{1_{q}},
\end{equation}
where $\ket{\{0_q\}}$ is the multi-mode cavity vacuum, $\ket{e_i}$ describes a single excitation in the $i$-th molecule with all the other dipoles in the ground state, and $\ket{1_q}$ describes a single photon in the $q$-th mode, with all the other modes in the vacuum state. There are no other restrictions on the  material and photonic coefficients $c_{i}^{(0)}$ and $c_{q}^{(1)}$ except for normalization $\sum_i |c_{i}^{(0)}|^2+\sum_q|c_{q}^{(1)}|^2=1$.

We derive evolution equations for the material and photonic wavepacket amplitudes $c_i^{(0)}$ and $c_q^{(1)}$ from the single-excitation ansatz  (\ref{eq:ansatz SM}), directly from the non-Hermitian Schrödinger equation $(d/dt)\ket{\psi^{1}}= -i\hat{\mathcal{H}}_{\rm eff} \ket{\psi^{(1)}}$, with $\hat{\mathcal{H}}_{\rm eff}$ given by Eq. (\ref{eq:Heff}). Rewriting the state amplitudes as $c_j^{(0)}= \tilde c_j^{(0)}\,{\rm exp}[-i \tilde \omega_0 t]$ and $c_q^{(1)}=\tilde c_q^{(1)} \,{\rm exp}[-i \tilde \omega_0  t]$, where $\tilde \omega_0=\omega_0-i\kappa_0/2$ is the complex frequency of the free $q=0$ mode. In this new  $q=0$ frame,  we obtain the following system of $(M+N)$ coupled equations for the amplitudes 
\begin{eqnarray}
\frac{d}{dt}\tilde c_j^{(0)} &=& (i\delta_j  + \Delta\gamma_j/2)\tilde  c_j^{(0)} - i \sum_q g_{jq} \tilde c_q^{(1)}\label{eq:cj}\\
\frac{d}{dt}\tilde c_q^{(1)} &=& -(i\Delta_q  + \Delta\kappa_q/2)\tilde  c_q^{(1)} - i \sum_j g_{jq}^* \tilde c_j^{(0)}\label{eq:cq},
\end{eqnarray}
where $\delta_j = \omega_0 - \omega_j$ is the detuning of the reference ($q=0$) mode relative to the $i$-th dipole frequency,  $ \omega_q= \omega_0+\Delta_q$ is the frequency of the $q$-th higher ($\Delta_q>0$) or lower order mode ($\Delta_q<0$) relative to $q=0$. $\Delta \gamma_j= \kappa_0-\gamma_j$ is the decay mismatch between dipoles and the reference mode, and $\Delta\kappa_q=\kappa_q-\kappa_0$ is decay mismatch relative to $q=0$.

We derive a reduced set of coupled equations of motion for the $N$ excited state amplitudes $\tilde c_j^{(0)}$ and the single-photon amplitude of the reference mode $\tilde c_0{(1)}$, under the assumption that modes that only the reference mode $q=0$ interacts resonantly with the dipolar ensemble, but higher and lower order modes ($q\neq 0$) are sufficiently detuned from the dipole frequencies $\omega_j$ to prevent significantly exchange of energy and coherence between light and matter. In this case, the off-resonant modes simply follow adiabatically the dipole polarization to lowest order in $g_{jq}$. The stationary off-resonant mode amplitude is obtained from Eq. (\ref{eq:cq}) to give
\begin{equation}\label{eq:cq adiabatic}
\tilde c_{q\neq 0}^{(1)} = -i\sum_{i=1}^N\frac{g_{iq}^* \,\tilde c_i^{(0)}}{i\Delta_q + \Delta\kappa_q/2}.
\end{equation}
Separating Eqs. (\ref{eq:cj}) and (\ref{eq:cq}) into resonant ($q=0$) and non-resonant mode contributions, and inserting the adiabatic solution in Eq. (\ref{eq:cq adiabatic}) for the lower and higher order modes, we obtain a reduced set of ($1+N$) equations of motion of the form
\begin{eqnarray}
\frac{d}{dt}\tilde c_j^{(0)} &=& i(\delta_j-\Gamma_j^{''})\tilde c_j^{(0)}  + (\Delta\gamma_j/2-\Gamma_j')\tilde  c_j^{(0)}-ig_{j0}\tilde c_0^{(1)}-\sum_{i\neq j}(J_{ij}'+iJ_{ij}^{''})\tilde c_i^{(0)}\label{eq:reduced cj}\\
\frac{d}{dt}\tilde c_0^{(1)} &=& -i\sum_i g_{i0}^* \,\tilde c_i^{(0)}\label{eq:reduced c0}
\end{eqnarray}
where we introduced the effective one-body and two-body decay rates,
\begin{eqnarray}\label{eq:effective decays}
\Gamma_j' &=& -\sum_{q\neq 0} |g_{jq}|^2\frac{ \Delta\kappa_q/2}{\Delta_q^2+( \Delta\kappa_q/2)^2}, \\
J_{ij}' &=& -\sum_{q\neq 0} g_{iq}^*g_{jq} \frac{ \Delta\kappa_q/2}{\Delta_q^2+( \Delta\kappa_q/2)^2},
\end{eqnarray}
and effective frequency shifts given by
\begin{eqnarray}\label{eq:effective frequencies}
\Gamma_j^{''} &=& -\sum_{q\neq 0} |g_{jq}|^2\frac{\Delta_q}{\Delta_q^2+(\Delta \kappa_q/2)^2}, \\
J_{ij}'' &=& -\sum_{q\neq 0} g_{iq}^*g_{jq} \frac{\Delta_q/2}{\Delta_q^2+(\Delta \kappa_q/2)^2}.
\end{eqnarray}
In these expressions, the sum-over-modes exclude the reference $q=0$ cavity resonance. The reduced equations of motion (\ref{eq:reduced cj}) and (\ref{eq:reduced c0}) can accurately describe strong coupling between the $N$ oscillators and the near-resonant $q=0$ mode, and improves over previous treatments in the literature by including the effect of far-detuned higher and lower order modes self-consistently. The quasi-static solution for non-resonant modes in Eq. (\ref{eq:cq adiabatic}) is accurate for modes that do not significantly admix with dipole transitions, which requires $|\Delta_q|>\sqrt{\sum_j|g_{jq}|^2}$ for each  $q\neq 0$.

In the idealized Dicke regime where the $N$ dipoles have equal transition frequencies ($\omega_i=\omega_e$), Rabi couplings ($g_{iq}=g_q$) and relaxation rates ($\gamma_i=\gamma$), the reduced dynamical equations (\ref{eq:reduced cj}) and (\ref{eq:reduced c0}) further simplify by writing $\tilde c_i = \beta/\sqrt{N}$ and $\tilde c_0= \alpha_0$. The evolution equation for the collective dipole coherence $\tilde X = \sum_i \tilde c_i^{(0)}= \sqrt{N}\beta$, can be obtained by summing Eq. (\ref{eq:reduced cj}) over all dipoles to obtain two coupled equations for $\alpha_0$ and $\beta$ that can be written in matrix form as
\begin{equation}\label{eq:dicke eom}
\left(\begin{array}{c}
\dot\alpha_0\\
\dot\beta
\end{array}\right)
= \left( \begin{array}{cc}
0 & -i\sqrt{N}g_0\\
-i\sqrt{N}g_0& i \delta_N - \Delta\Gamma_N
\end{array}\right)\left(\begin{array}{c}
\alpha_0\\
\beta
\end{array}\right), 
\end{equation}
with
\begin{eqnarray}
\delta_N&=& \delta_0-\Gamma{''}-NJ{''}\\
\Delta\Gamma_N &=&  -\Delta\gamma/2 +\Gamma' + NJ{'}.
\end{eqnarray}
This is Eq. (\ref{eq:adiabatic Ham}) in the main text. $\delta_0 = \omega_0-\omega_e$, $\Delta\gamma = \kappa_0-\gamma$, and $g_0$ refer to properties of the near-resonant $q=0$ mode. 

\section{Approximate Rabi splitting and polariton bandwidths in low-finesse three-mode cavities}
\label{app:low finesse}

In this section we derive Eqs. (\ref{eq:polariton splitting}) and (\ref{eq:polariton bandwidths}) in the main text. \\

We consider a three-mode cavity model with a central $q=0$ mode at $\omega_0$ tuned near resonance with a homogeneous ensemble of $N$ dipole transitions at $\omega_e$. The dipoles also couple to the lower-order mode $q=-1$ and a higher-order mode $q=+1$, with frequencies $\omega_{\pm1}=\omega_0\pm\Delta_{\pm 1}$, respectively. The evolution equations (\ref{eq:cj})-(\ref{eq:cq}) reduce in this case to a (3+1) system for the collective coherence $\tilde X = \sum_j\tilde c_j^{0}=\sqrt{N}\beta$, ($\delta_j=\delta_0$, $\Delta\gamma_j=\Delta \gamma$, $g_{jq}=g_q$) and photon amplitudes $\tilde c_q^{(1)}\equiv \alpha_q$, with $q=\{-1,0,1\}$. The resulting system of equations can be written in matrix form as $\dot{\mathbf{x}}= -i\mathbf{M}\, \mathbf{x}$, with $\mathbf{x}=[\alpha_{+1},\alpha_0,\alpha_{-1},\beta]^{\rm T}$ and $\mathbf{M}$ written in  arrowhead form as \footnote{Note that the arrowhead interaction matrix is not the only possibility, other alternatives have also been explored~\cite{Richter_APL_2015_107_231104}. For our purposes here the arrowhead approach provides a convenient starting point to explore the physics we wish to investigate. Future work will need to explore the consequences of other choices.}, 

\begin{equation}\label{eq:three-mode matrix}
\mathbf{M}= \left( \begin{array}{cccc}
\Delta_{-1}-i\Delta\kappa_{-1}/2 &0&0& \Omega_{-1}\\
0 &0&0& \Omega_{0}\\
0 &0& \Delta_{+1}-i\Delta\kappa_{+1}/2& \Omega_{+1}\\
\Omega_{-1} &\Omega_{0}&\Omega_{+1}& -\delta +i\Delta\gamma/2\\
\end{array}\right)\equiv \left( \begin{array}{cccc}
d_{-1} &0&0& e_{-1}\\
0 &0&0& e_{0}\\
0 &0& d_{+1}& e_{+1}\\
e_{-1} &e_{0}&e_{+1}& p\\
\end{array}\right),
\end{equation}

\noindent where $\Omega_q=\sqrt{N}g_q$ (with $g_q=g_q^*$), $\delta=\omega_0-\omega_e$, $\Delta\gamma=\kappa_0-\gamma$, $\Delta\kappa_q=\kappa_q-\kappa_0$,and $\Delta_q=\omega_q-\omega_0$. We introduced a simplified notation in the second equality. The complex eigenvalues $\lambda=E+i\Gamma/2$ of $\mathbf{M}$ are  roots of the characteristic polynomial \cite{OLeary:1990}  
\begin{equation}\label{eq:characteristic poly}
\Phi(\lambda)\equiv p-\lambda +\frac{e_0^2}{\lambda}- \frac{e_{-1}^2}{d_{-1}- \lambda} -\frac{e_{+1}^2}{d_{+1}- \lambda}.
\end{equation}
We are interested in corrections to the polariton splitting $\Omega_R \equiv {\rm Re}[\lambda_{\rm UP}]-{\rm Re}[\lambda_{\rm LP}]$ and bandwidths $\Gamma_{\rm LP}=2{\rm Im}[\lambda_{\rm LP}]$, $\Gamma_{\rm UP}=2{\rm Im}[\lambda_{\rm UP}]$, from their values in a single-mode scenario where $q=0$ is resonant with the dipoles ($\delta\approx 0$). Corrections are due to the presence of the adjacent $q=\pm 1$ modes. We thus solve for $x$ in $\Phi(\lambda_\pm+x)=0$, where $\lambda_\pm$ denotes the pair of coupled single-mode solutions to Eq. (\ref{eq:characteristic poly}) obtained for $\Omega_{\pm 1}=0$, giving $\lambda_\pm(p-\lambda_\pm)+e_0^2=0$. For $\Delta \gamma=\delta=0$ ($p=0$), the single-mode LP and UP solutions are $\lambda_{\pm}=\pm  \Omega_0$, giving the bare splitting  $\Omega_R=2\Omega_0$ and bandwidths $\Gamma_{\rm LP}=\Gamma_{\rm UP}=\kappa$. 

Linearizing the polynomial $\Phi(\lambda_{\pm}+x)=0$ around $x=0$ gives the general solution,
\begin{equation}
x_{\pm} = \frac{-\lambda_{\pm}[e_{-1}^2d_{+1}+e_{+1}^2d_{-1}-(e_{-1}^2+e_{+1}^2)\lambda_{\pm}]}{P(\lambda_\pm)}, 
\end{equation}
with 
\begin{eqnarray}
    P(\lambda_\pm)&=&d_{+1}(e_0^2+e_{-1}^2)+d_{-1}(e_0^2+e_{+1}^2)-d_{-1}d_{+1}p-2(e_0^2+e_{-1}^2+e_{+1}^2-d_{+1}p-d_{-1}p-d_{-1}d_{+1})\lambda_\pm\nonumber\\
    &&-3(d_{-1}+d_{+1}+p)\lambda_\pm^2+4\lambda_\pm^3.
\end{eqnarray}
The solution $\lambda_{\rm LP}=\lambda_-+x_-$ gives the LP frequency and decay rate and $\lambda_{\rm UP}= \lambda_++x_+$ gives the UP properties. The denominator $P(\lambda_\pm)$ can be reduced by setting $p=0$, which holds for a resonant single-mode scenario ($\delta_0=0$) without decay mismatch ($\Delta\gamma=0$), giving
\begin{equation}\label{eq:xLPUP}
x_\pm= -\lambda_\pm\times 
\frac{d_{+1}e_{-1}^2+d_{-1}e_{+1}^2-\lambda_{\pm}(e_{-1}^2+e_{+1}^2)}{d_{+1}(e_0^2+e_{-1}^2)+d_{-1}(e_0^2+e_{+1}^2)-2(e_0^2+e_{+1}^2+e_{-1}^2-d_{-1}d_{+1})\lambda_{\pm}-3(d_{-1}+d_{+1})\lambda_\pm^2+4\lambda_\pm^3}.\nonumber\\
\end{equation}
For the special case where there are no mode-order variations of the Rabi coupling and cavity mode bandwidth, i.e., $\Omega_{q}=\Omega$ and $\kappa_q=\kappa$ for all $q$, and assuming modes that are equally spaced ($\Delta_\pm=\pm\Delta$), we have $d_{+1}e_{-1}^2+d_{-1}e_{+1}^2=+\Delta\Omega^2-\Delta\Omega^2=0$ in Eq. (\ref{eq:xLPUP}) and corrections to the LP and UP eigenvalues become purely real
\begin{equation}
    x_\pm=-\lambda_{\pm}\,  \frac{\Omega^2}{3\Omega^2+\Delta^2-2\lambda_\pm^2}=\mp \Omega\left(\frac{\Omega^2}{\Omega^2+\Delta^2}\right),
\end{equation}
with $\Delta>\Omega$ and $\lambda_\pm=\pm \Omega$. Therefore, without mode-order variations of the cavity parameters, level pushing from the higher-order and lower-order modes gives the reduced Rabi splitting \begin{equation}
\Omega_R=2\Omega+x_+-x_-=2\Omega\left(\frac{\Delta^2}{\Delta^2+\Omega^2}\right),
\end{equation}
but there no changes to the polariton bandwidths relative to the single-mode scenario.

The simplest ideal scenario with corrections to the LP and UP bandwidths is one where the Rabi couplings do not depend on mode order ($\Omega_p=\Omega$), but cavity modes can differ in bandwidth ($\Delta\kappa_q\neq 0$). From Eq. (\ref{eq:xLPUP}), neglecting quadratic terms in $\Delta\kappa_{\pm 1}$ relative to $\Delta^2+\Omega^2$, we obtain in this case, 
\begin{equation}\label{eq:x real}
    {\rm Re}[x_\pm] \approx \mp\Omega\,\frac{\Omega^2(\Omega^2+\Delta^2)^2}{(\Omega^2+\Delta^2)^2+(\Delta(\Delta\kappa_{-1}-\Delta\kappa_{+1})\mp \Omega(\Delta\kappa_{-1}+\Delta\kappa_{+1})/2)^2},
\end{equation}
and,
\begin{equation}\label{eq:x imag}
    2{\rm Im}[x_\pm] = \mp\Omega\,\frac{\Omega(\Delta\kappa_{-1}+\Delta\kappa_{+1})(\Omega^2+\Delta^2)/2\pm \Omega^2[\Delta(\Delta\kappa_{-1}-\Delta\kappa_{+1})\mp \Omega(\Delta\kappa_{-1}+\Delta\kappa_{+1})/2]}{(\Omega^2+\Delta^2)^2+(\Delta(\Delta\kappa_{-1}-\Delta\kappa_{+1})\mp \Omega(\Delta\kappa_{-1}+\Delta\kappa_{+1})/2)^2}.
\end{equation}
From these expressions we may obtain the lower and upper polariton energies $E_\pm= \pm \Omega+{\rm Re}[x_\pm]$ and decay rates $\Gamma_{\pm}=\kappa+2{\rm Im}[x_\pm]$. For  linear changes in mode bandwidth of the form $\Delta\kappa_q=q\zeta$, with $\zeta$ positive or negative, the LP and UP energies can then be written as,
\begin{equation}\label{eq:energies linear change}
E_\pm =\pm \Omega\left[1-\frac{\Omega^2(\Omega^2+\Delta^2)}{(\Omega^2+\Delta^2)^2+\Delta^2\zeta^2}\right],
\end{equation}
from where Eq. (\ref{eq:polariton splitting}) in the main text is obtained. From Eq. (\ref{eq:x imag}), the polariton bandwidths for linear bandwidth changes with mode order are thus
\begin{equation}\label{eq:bandwidths linear change}
 \Gamma_\pm=\kappa \pm 2\Omega \frac{\Omega^2\Delta\zeta}{(\Omega^2+\Delta^2)^2+\Delta^2\zeta^2}, 
\end{equation}
which is Eq. (\ref{eq:polariton bandwidths}) in the main text.

\end{widetext}

\vspace{2cm}

\bibliographystyle{unsrt}
\bibliography{finesse_philtrans.bib}

\end{document}